\documentclass[a4paper,english,aps]{revtex4}%
\usepackage[T1]{fontenc}
\usepackage[latin9]{inputenc}
\usepackage{amsmath}
\usepackage{graphicx}
\usepackage{epsfig}
\usepackage{esint}
\usepackage{babel}%
\usepackage{amsfonts}%
\usepackage{amssymb}
\makeatletter
\providecommand{\tabularnewline}{\\}
\makeatother
\begin{document}
\preprint{niekompletny szkic}
\title{Covariant Non-local Chiral Quark Model and Pion-photon Transition Distribution Amplitudes}
\author{Piotr Kotko}
\email{kotko@th.if.uj.edu.pl}
\author{Michal Praszalowicz}
\email{michal@if.uj.edu.pl}
\affiliation{M. Smoluchowski Institute of Physics, Jagiellonian University, ul. Reymonta 4,
30-059 Krak{\'{o}}w, Poland}

\begin{abstract}
Using non-local chiral quark model with simple pole ansatz for mass dependence
on momentum and non-local currents satisfying Ward-Takahashi identities, we
calculate pion to photon transition distribution amplitudes and relevant form
factors. For vector amplitude we recover correct normalization fixed by the
axial anomaly. We find that due to the non-locality in the vector current, the
value of axial form factor at zero momentum transfer is lowered with respect
to the vector form factor. Such behaviour -- consistent with experiment -- is
not seen in the local models. Where possible we compare our results to the
experimental data.

\end{abstract}
\maketitle

\section{Introduction}

Transition Distribution Amplitudes (TDA) were firstly introduced in ref.
\cite{Pire} as objects parametrizing soft part of the amplitudes for
hadron-antihadron annihilation $\bar{H}H\rightarrow\gamma^{*}\gamma$ or
backward Compton scattering $\gamma^{*}H\rightarrow H\gamma$, where the hard
scale is provided by high virtuality of one of the photons. In some sense
TDA's are hybrids of the ordinary Distribution Amplitudes (DA) and Generalized
Parton Distributions (GPD) (see \cite{Belitsky} for a review). However, if
one restricts oneself to the mesonic case, they are more similar to GPD's -
the difference is that we deal with matrix elements which are non-diagonal not
only in momenta but in the physical states as well. Nevertheless, from
kinematical point of view, both are almost identical and therefore we can use
similar variables as skewedness for example. In practice we consider two kinds
of TDA's: vector and axial, depending on nature of the bilocal quark-antiquark
operator sandwiched between photon and meson states. For experimental issues of
Transition Distribution Amplitudes see ref. \cite{Pire2}.

In order to avoid complexity of bound state physics we limit ourselves to
pions only. Pions are Goldstone bosons of broken SU(2) chiral symmetry and
their properties are to large extent determined by the symmetry (breaking)
alone rather than by the complex phenomenon of confinement. After such
simplification, we can use the non-local semi-bosonized Nambu-Jona-Lasinio
(NJL) model to get an insight into TDA's. On the other hand, since Transition
Distribution Amplitudes are related to anomalous diagrams, they can serve as a
demanding tester of the non-local models. We shall come back to this point
later in this paper.

First estimates of TDA's were made in refs. \cite{Tiburzi,BronTDA,Noguera}.
The first one \cite{Tiburzi} was based on general QCD symmetries supported 
by simple quark model, while the second \cite{BronTDA} and the third one 
\cite{Noguera} made use of Spectral Quark Model (SQM)
\cite{SQM} and Pauli-Villars regulated NJL, respectively. The calculations
in SQM respect all QCD symmetries by definition (i.e. Lorentz invariance, Ward
identities, anomalies etc.).  In Pauli-Villars regulated NJL model the presence 
of the finite regulator \cite{Noguera} gives normalization for the vector amplitude
which is not
consistent with axial anomaly. In \cite{KotkoMP_TDA} we calculated TDA's
numerically in the non-local chiral quark model, however, since we have used
"naive" currents we also broke the normalization condition of vector TDA.

In the present paper we extend our calculations from ref. \cite{KotkoMP_TDA}
in twofold way: we use the full non-local vertices and obtain not only
numerical but also analytical results for some of the TDA's. As a result we
recover correct normalization conditions fixed by axial anomaly and give
predictions for axial form factor which is not restricted by anomaly. Our
results are here in qualitative agreement with experimental data. Since we
revised our old calculations using different method we traced the mistake in
the vector amplitude, which however does not change qualitative results.

The paper is organized as follows. In Section \ref{model} we review the
non-local chiral quark model and give the prescription for modified currents.
Next, in Section \ref{sec:definitions}, we recall the definitions of TDA's and
relevant sum rules. Section \ref{sec:Results} contains our results with
special emphasis on the role of the new pieces coming from the non-local parts
of the currents. In Section \ref{sec:formfactors} we investigate the relevant
form factors and finally we summarize our results in Section \ref{sec:summary}%
. Technical details are given in the Appendices.

\section{Non-local covariant chiral quark model}

\label{model}

In the following we use the non-local semi-bosonized Namu-Jona-Lasinio model.
It is based on the quark-pion interaction in the following form \cite{Diakonov1}%
\begin{equation}
S_{\mathsf{Int}}=\int\frac{d^{4}k\,d^{4}l}{(2\pi)^{8}}\bar{\psi}%
(k)\sqrt{M\left(  k\right)  }U^{\gamma_{5}}(k-l)\sqrt{M(l)}\psi
(l),\label{eq:action}%
\end{equation}
where $M\left(  k\right)  $ is dynamical quark mass appearing due to the
spontaneous chiral symmetry breaking. Meson field $U^{\gamma_{5}}$ is given in
terms of the pion field as%
\begin{equation}
U^{\gamma_{5}}\left(  x\right)  =\exp\left\{  \frac{i}{F_{\pi}}\tau^{a}\pi
^{a}\left(  x\right)  \gamma_{5}\right\}  ,\label{eq:Ufield}%
\end{equation}
where $F_{\pi}=93\,\mathrm{MeV}$ is the pion decay constant.

One defines mass dependence on momentum as%
\begin{equation}
M\left(  k\right)  =M\,F^{2}\left(  k\right) \label{eq:mass}%
\end{equation}
where form factor $F\left(  k\right)  $ should vanish for $k\rightarrow\infty$
and is chosen to satisfy $F\left(  0\right)  =1$. Expression for $F\left(
k\right)  $ was obtained analytically in Euclidean space from the instanton
model of the QCD vacuum and it is highly non trivial \cite{Diakonov1}.
Therefore here we use the Minkowski form proposed in \cite{MP_pion}
\begin{equation}
F(k)=\left(  \frac{-\Lambda_{n}^{2}}{k^{2}-\Lambda_{n}^{2}+i\epsilon}\right)
^{n},\label{Fkdef}%
\end{equation}
which reproduces reasonably well the instanton result when continued to
Euclidean space. However, it can be also used directly in the Minkowski space.
Integer parameter $n$ defines a family of models and allows to analyze a
dependence of our results on the shape of $F\left(  k\right)  $. It was argued
in ref. \cite{Dorokhov_VAV} that $F\left(  k\right)  $ should vanish
exponentially for large momenta in order to be consistent with OPE. This is
another reason for introducing $n$ parameter, which allows to control high
momentum behavior. The method of fixing value of $\Lambda_{n}$ is described below.

It is a well known fact that due to the mass dependence on momentum the
standard vector and axial currents are not conserved and Ward-Takahashi (WT)
identities are not satisfied. Although it was argued in ref.
\cite{KotkoMP_TDA} that this violation is not very large and almost does not
affect shapes of TDA's, it certainly affects overall normalizations. It is
therefore necessary to resolve the problem of normalization if one wants to use
TDA's calculated in the non-local model for phenomenological estimates.

There are several ways of constructing conserved currents in presence of
non-local interactions see refs.
\cite{PagelsStokar,BallChiu,Holdom,Birse,Frank}. 
However, it should be recalled at this point that none of them is
unique, because current conservation fixes only the longitudinal part of a
given vertex while the transverse one has to be modeled. In this paper we use
the simplest {}\textquotedblleft minimal\textquotedblright\ vertices
satisfying WT identities. For vector current $j^{\mu}=\bar{\psi}\gamma^{\mu
}\psi$ we replace $\gamma^{\mu}$ by the following non-local vertex%
\begin{equation}
\Gamma^{\mu}\left(  k,p\right)  =\gamma^{\mu}+g^{\mu}\left(  k,p\right)
,\label{eq:vector_vertex}%
\end{equation}
where the non-local addition reads%
\begin{equation}
g^{\mu}\left(  k,p\right)  =-\frac{k^{\mu}+p^{\mu}}{k^{2}-p^{2}}\left(
M\left(  k\right)  -M\left(  p\right)  \right)  .\label{eq:vector_vertex_1}%
\end{equation}
This form of non-locality does not introduce singularities as required by
general prosperities of vector vertices \cite{BallChiu}.

Axial current $j_{5}^{\mu}=\bar{\psi}\gamma^{\mu}\gamma_{5}\psi$ is made
conserved by replacing $\gamma^{\mu}\gamma_{5}$ by%
\begin{equation}
\Gamma_{5}^{\mu}\left(  k,p\right)  =\gamma^{\mu}\gamma_{5}+g_{5}^{\mu}\left(
k,p\right)  ,\label{eq:axial_vertex}%
\end{equation}
with%
\begin{equation}
g_{5}^{\mu}\left(  k,p\right)  =\frac{p^{\mu}-k^{\mu}}{\left(  p-k\right)
^{2}}\left(  M\left(  k\right)  +M\left(  p\right)  \right)  \gamma
_{5}.\label{eq:axial_vertex1}%
\end{equation}
In contrary to the vector vertex, the axial one contains physical singularity
corresponding to the pion.

For a model given by $n$ we fix $\Lambda_{n}$ using the Birse-Bowler
\cite{Birse} formula for the pion decay constant $F_{\pi}$:
\begin{equation}
F_{\pi}^{2}=\frac{N_{c}}{4\pi^{2}}\int_{0}^{\infty}dk_{\mathrm{E}}^{2}%
\frac{M^{2}\left(  k_{\mathrm{E}}\right)  -k_{\mathrm{E}}^{2}M\left(
k_{\mathrm{E}}\right)  M^{\prime}\left(  k_{\mathrm{E}}\right)  +k_{\mathrm{E}%
}^{4}M^{\prime}\left(  k_{\mathrm{E}}\right)  ^{2}}{\left(  k_{\mathrm{E}}%
^{2}+M^{2}\left(  k_{\mathrm{E}}\right)  \right)  ^{2}}\label{eq:Fpi}%
\end{equation}
with $F_{\pi}=93\,\mathrm{MeV}$. Using \eqref{Fkdef} this expression can be
calculated analytically - see the Appendix of \cite{KotkoMP_foton}. In Table
\ref{tab:model_param} we present $\Lambda_{n}$ for given values of constituent
quark mass $M$ and power $n$. Let us recall at this point that there exists
another expressions for $F_{\pi}$, namely the Pagels-Stokar formula
\cite{PagelsStokar}. As shown explicitly in \emph{e.g. } \cite{BzdakMP} it
accommodates a non-local pion-quark interaction, however it corresponds to the
naive axial current $j_{5}^{\mu}$ which, as mentioned above, does not exhibit
PCAC. On the contrary the Birse-Bowler formula \cite{Birse} takes into
account the non-local axial current (more precisely its derivative, which is
determined unambiguously). See ref. \cite{BzdakMP} for discussion.

It is important to note that $\Lambda_{n}$ \emph{does not correspond} to the
QCD scale characteristic for the present model. Indeed function $F(k)$ does
not change much for different $(n,\Lambda_{n})$ except for the high momentum
part. Therefore, as pointed out in \cite{MP_Conden,KotkoMP_TDA}, the precise
definition of the scale can be done only within QCD, while within an effective
model one can only estimate the order of magnitude. For the instanton model
the characteristic scale is about $600\,\mathrm{MeV}$ \cite{Diakonov1}. A
somewhat more detailed discussion of this issue can be found in ref. \cite{MP_Conden}.

\begin{table}[ptb]
\begin{centering}
\begin{tabular}{|c|c|}
\hline
\multicolumn{2}{|c|}{$M=225\,\mathrm{MeV}$}\tabularnewline
\hline
\hline
$n=1$ & $\Lambda=1641\,\mathrm{MeV}$\tabularnewline
\hline
$n=5$ & $\Lambda=3823\,\mathrm{MeV}$\tabularnewline
\hline
\hline
\multicolumn{2}{|c|}{$M=350\,\mathrm{MeV}$}\tabularnewline
\hline
\hline
$n=1$ & $\Lambda=836\,\mathrm{MeV}$\tabularnewline
\hline
$n=5$ & $\Lambda=1970\,\mathrm{MeV}$\tabularnewline
\hline
\hline
\multicolumn{2}{|c|}{$M=400\,\mathrm{MeV}$}\tabularnewline
\hline
\hline
$n=1$ & $\Lambda=721\,\mathrm{MeV}$\tabularnewline
\hline
$n=5$ & $\Lambda=1704\,\mathrm{MeV}$\tabularnewline
\hline
\end{tabular}
\par\end{centering}
\caption{Numerical values of the model parameters obtained using Birse-Bowler
formula \eqref{eq:Fpi} for the pion decay constant $F_{\pi}$. }%
\label{tab:model_param}%
\end{table}

The non-local model with momentum dependent quark mass given by \eqref{Fkdef}
was applied in the past to several low energy quantities (pion DA
\cite{MP_pion}, two pion generalized DA \cite{MP_GDA}, quark and gluon
condensates \cite{MP_Conden}), using however naive vector and axial currents.
Later in ref. \cite{BzdakMP}, pion DA with conserved axial current was
estimated using PCAC. Recently a set of photon DA's up to twist-4 was obtained
using currents satisfying WT identities \cite{KotkoMP_foton}.

Similar models with however different form of $F\left(  k\right)  $ in
\eqref{eq:mass} were also considered in the literature. In ref.
\cite{Birse} exponential form of $F\left(  k\right)  $ was used. In
refs.
\cite{PetrovPolyakov_pion_foton,Dorokhov1,Dorokhov2,Dorokhov3,Dorokhov_pion,Bron_foton} 
various distribution amplitudes and correlators have been studied in non-local models similar
to ours. Pion and kaon DA were calculated also in \cite{Kim}.

\section{Kinematics and definitions}

\label{sec:definitions}

Relevant kinematics is very close to the one used in GPD formalism. We
consider pion with momentum $P_{1}$ and real photon carrying momentum $P_{2}$.
We shall work in the chiral limit, therefore both $P_{1}^{2}=0$ and $P_{2}%
^{2}=0$. We define momentum transfer $q^{\mu}=P_{2}^{\mu}-P_{1}^{\mu} $ and
momentum transfer squared $t=q^{2}$. Two light-like directions are defined by
null vectors $n=\left(  1,0,0,-1\right)  $ and $\tilde{n}=\left(
1,0,0,1\right)  $. Decomposition of any vector $v^{\mu}$ in this basis reads%
\begin{equation}
v^{\mu}=v^{+}\frac{\tilde{n}^{\mu}}{2}+v^{-}\frac{n^{\mu}}{2}+v_{T}^{\mu
},\label{eq:lightcoord1}%
\end{equation}
while the scalar product%
\begin{equation}
u\cdot v=\frac{1}{2}u^{+}v^{-}+\frac{1}{2}u^{-}v^{+}-\vec{u}_{T}\cdot\vec
{v}_{T},\label{eq:lightcoord2}%
\end{equation}
where arrows denote Euclidean two-vectors. Introducing average momentum
$p=\frac{1}{2}\left(  P_{1}+P_{2}\right)  $ we can define so called skewedness
variable%
\begin{equation}
\xi=-\frac{q^{+}}{2p^{+}}.\label{eq:skew}%
\end{equation}
For massless pions $-1<\xi<1$. Switching to the frame where the average
momentum does not have transverse part, \textit{i.e.}%
\begin{equation}
p^{\mu}=p^{+}\frac{\tilde{n}^{\mu}}{2}+\frac{p^{2}}{p^{+}}\,\frac{n^{\mu}}%
{2}\label{eq:p_av}%
\end{equation}
we can write the following parametrizations of the pion and photon momenta%
\begin{equation}
P_{1}^{\mu}=\left(  1+\xi\right)  p^{+}\frac{\tilde{n}^{\mu}}{2}+\left(
1-\xi\right)  \frac{p^{2}}{p^{+}}\,\frac{n^{\mu}}{2}-\frac{1}{2}q_{T}^{\mu
}\label{eq:P1}%
\end{equation}%
\begin{equation}
P_{2}^{\mu}=\left(  1-\xi\right)  p^{+}\frac{\tilde{n}^{\mu}}{2}+\left(
1+\xi\right)  \frac{p^{2}}{p^{+}}\,\frac{n^{\mu}}{2}+\frac{1}{2}q_{T}^{\mu
}.\label{eq:P2}%
\end{equation}
Notice that $p^{2}=-t/4$. We denote photon polarization vector by
$\varepsilon$. It satisfies condition $\varepsilon\cdot P_{2}=0$.

We can now define TDA's. Vector Transition Distribution Amplitude (VTDA)
$V\left(  X,\xi,t\right)  $ is defined as%
\begin{equation}
\int\frac{d\lambda}{2\pi}e^{i\lambda Xp^{+}}\left\langle \gamma\left(
P_{2},\varepsilon\right)  \left\vert \overline{d}\left(  -\frac{\lambda}%
{2}n\right)  \Gamma^{\mu}u\left(  \frac{\lambda}{2}n\right)  \right\vert
\pi^{+}\left(  P_{1}\right)  \right\rangle 
=\frac{i^{2}e}{2\sqrt{2}F_{\pi}p^{+}}\varepsilon^{\mu\nu\alpha\beta
}\varepsilon_{\nu}^{\ast}p_{\alpha}q_{\beta}V\left(  X,\xi,t\right)
.\label{eq:def_VTDA}%
\end{equation}
In the axial channel the Axial Transition Distribution Amplitude (ATDA)
$A\left(  X,\xi,t\right)  $ is defined as:%
\begin{multline}
\int\frac{d\lambda}{2\pi}e^{i\lambda Xp^{+}}\left\langle \gamma\left(
P_{2},\varepsilon\right)  \left\vert \overline{d}\left(  -\frac{\lambda}%
{2}n\right)  \Gamma_{5}^{\mu}u\left(  \frac{\lambda}{2}n\right)  \right\vert
\pi^{+}\left(  P_{1}\right)  \right\rangle 
=\frac{ie}{2\sqrt{2}F_{\pi}p^{+}}P_{2}^{\mu}q\cdot\varepsilon^{\ast}A\left(
X,\xi,t\right) \\
+q^{\mu}q\cdot\varepsilon^{\ast}\frac{ie2\sqrt{2}F_{\pi}\mathrm{sign}\left(
\xi\right)  }{t}\phi_{\pi}\left(  \frac{X+\xi}{2\xi}\right)  +\ldots
.\label{eq:defATDA}%
\end{multline}
Notice that we have explicitly written down the contribution containing
massless pole $1/t$ coming from the fact that the pion can couple directly to
the axial current. This part is connected with the pion DA $\phi_{\pi}\left(
u\right)  $. Dots stand for the remaining parts which vanish when contracted
with $n^{\mu}$ \cite{Tiburzi}.

There exist the following sum rules for TDA's%
\begin{equation}
{\displaystyle\int\limits_{-1}^{1}}dX\,\left\{
\begin{array}
[c]{c}%
V\left(  X,\xi,t\right) \\
A\left(  X,\xi,t\right)
\end{array}
\right.  =\frac{2\sqrt{2}F_{\pi}}{m_{\pi}}\left\{
\begin{array}
[c]{c}%
F_{V}\left(  t\right) \\
F_{A}\left(  t\right)
\end{array}
\right.  =2\sqrt{2}F_{\pi}\left\{
\begin{array}
[c]{c}%
F_{V}^{\chi}\left(  t\right) \\
F_{A}^{\chi}\left(  t\right)
\end{array}
\right.  ,\label{eq:sumrules}%
\end{equation}
where $F_{V}\left(  t\right)  $ and $F_{A}\left(  t\right)  $ are vector and
axial form factors respectively. Quantity $F_{V,A}^{\chi}\left(  t\right)
=F_{V,A}\left(  t\right)  /m_{\pi}$ is introduced because we work in the
chiral limit ($m_{\pi}$ is the mass of the pion). Moreover the transition form
factor for the process $\pi^{0}\rightarrow\gamma^{\ast}\gamma$ is related to
the vector form factor by the formula%
\begin{equation}
F_{\pi\gamma}\left(  t\right)  =\sqrt{2}F_{V}^{\chi}\left(  t\right)
.\label{eq:transff}%
\end{equation}
The value of $F_{\pi\gamma}$ for $t=0$ is fixed by the axial anomaly and
equals%
\begin{equation}
F_{\pi\gamma}\left(  0\right)  =\frac{1}{4\pi^{2}F_{\pi}}\approx
0.272\,\mathrm{GeV}^{-1}.\label{eq:norm1}%
\end{equation}
There is no such a constraint for the axial form factor. See section
\ref{sec:formfactors} for the discussion.

\section{TDA's in covariant non-local model}

\label{sec:Results}

In this section we outline some points of the calculations within the model
introduced in the previous Section and show our results. More details are
given in the Appendix \ref{sub:App_Vector} and \ref{sub:App_Axial}.

\subsection{Vector TDA}

\label{sub:VTDA}

In the vector channel direct calculation of the matrix elements gives%
\begin{multline}
\int\frac{d\lambda}{2\pi}e^{i\lambda Xp^{+}}
\left\langle \gamma\left(  P_{2},\varepsilon\right)  \left\vert
\overline{d}\left(  -\frac{\lambda}{2}n\right)  \Gamma^{\mu}\, u\left(
\frac{\lambda}{2}n\right)  \right\vert \pi^{+}\left(  P_{1}\right)
\right\rangle \\
=-\frac{\sqrt{2}eMN_{c}}{F_{\pi}} \left(  Q_{d}\mathcal{M}_{V}^{\mu\nu
}\left(  X,\xi,t\right)  +Q_{u}\mathcal{M}_{V}^{\mu\nu}\left(  -X,\xi
,t\right)  \right)  \varepsilon_{\nu}^{\ast},\label{eq:3}%
\end{multline}
where $N_{c}$ is number of colors, $Q_{u},$ $Q_{d}$ are charges of the
pertinent quarks and%
\begin{multline}
\mathcal{M}_{V}^{\mu\nu}=\int\frac{d^{4}k}{\left(  2\pi\right)  ^{4}}%
\,\delta\left(  k^{+}-\left(  X-1\right)  p^{+}\right)  \times\nonumber\\
\mathrm{Tr}\left\{  S\left(  k+P_{2}\right)  \Gamma^{\mu}\left(
k+P_{2},k+P_{1}\right)  S\left(  k+P_{1}\right)  \Gamma_{5}\left(
k+P_{1},k\right)  S\left(  k\right)  \Gamma^{\nu}\left(  k+P_{1},k\right)
\right\} \label{eq:4}%
\end{multline}
with%
\begin{equation}
S\left(  p\right)  =\frac{1}{\not p-M\left(  p\right)  }.\label{eq:4a}%
\end{equation}
In the formula above quark-pion coupling reads%
\begin{equation}
\Gamma_{5}\left(  k,p\right)  =F\left(  k\right)  \gamma_{5}F\left(  p\right)
.\label{eq:pion-quark_coupling}%
\end{equation}

Notice that we have used the non-local vertex (\ref{eq:vector_vertex}) also in
the bilocal currents. This is necessary if one wants to maintain WT identities.

Amplitude $\mathcal{M}_{V}^{\mu\nu}$, due to the natural splitting of the
vector vertices to local and non-local parts, can be divided into four pieces%
\begin{equation}
\mathcal{M}_{V}^{\mu\nu}=\mathcal{M}_{V}^{\mu\nu,\,\left(  0\right)
}+\mathcal{M}_{V}^{\mu\nu,\,\left(  1\right)  }+\mathcal{M}_{V}^{\mu
\nu,\,\left(  2\right)  }+\mathcal{M}_{V}^{\mu\nu,\,\left(  3\right)
}.\label{eq:V1}%
\end{equation}
To list them we introduce shorthand notation
\begin{equation}
\hat{dk}=\frac{d^{4}k}{\left(  2\pi\right)  ^{4}}\,\delta\left(  k^{+}-\left(
X-1\right)  p^{+}\right)  .\label{eq:V1a}%
\end{equation}
Then%
\begin{equation}
\mathcal{M}_{V}^{\mu\nu,\,\left(  0\right)  }=\int\hat{dk}\,\mathrm{Tr}%
\left\{  S\left(  k+P_{2}\right)  \gamma^{\mu}S\left(  k+P_{1}\right)
\Gamma_{5}\left(  k+P_{1},k\right)  S\left(  k\right)  \gamma^{\nu}\right\}
\label{eq:V2}%
\end{equation}
corresponds to the quantities calculated in \cite{KotkoMP_TDA}. The new
pieces are
\begin{equation}
\mathcal{M}_{V}^{\mu\nu,\,\left(  1\right)  }=\int\hat{dk}\,\mathrm{Tr}%
\left\{  S\left(  k+P_{2}\right)  \gamma^{\mu}S\left(  k+P_{1}\right)
\Gamma_{5}\left(  k+P_{1},k\right)  S\left(  k\right)  g^{\nu}\left(
k+P_{1},k\right)  \right\}  ,\label{eq:V3}%
\end{equation}%
\begin{equation}
\mathcal{M}_{V}^{\mu\nu,\,\left(  2\right)  }=\int\hat{dk}\,\mathrm{Tr}%
\left\{  S\left(  k+P_{2}\right)  g^{\mu}\left(  k+P_{2},k+P_{1}\right)
S\left(  k+P_{1}\right)  \Gamma_{5}\left(  k+P_{1},k\right)  S\left(
k\right)  \gamma^{\nu}\right\}  .\label{eq:V4}%
\end{equation}
The remaining part of $\mathcal{M}_{V}^{\mu\nu}$ is zero, simply due to
vanishing of the Dirac trace, $\mathcal{M}_{V}^{\mu\nu,\,\left(  3\right)
}=0$.

Above decomposition, after projecting on the proper tensor structures, leads
straightforwardly to the expression for VTDA%
\begin{equation}
V\left(  X,\xi,t\right)  =V^{\left(  0\right)  }\left(  X,\xi,t\right)
+V^{\left(  1\right)  }\left(  X,\xi,t\right)  +V^{\left(  2\right)  }\left(
X,\xi,t\right)  .\label{eq:5}%
\end{equation}
The $V^{\left(  1\right)  }+V^{\left(  2\right)  }$ part is the addition that
is required in order to recover correct normalization. The explicit
expressions, together with some details needed to calculate $V^{\left(
0\right)  }$ analytically are given in the Appendix \ref{sub:App_Vector}.

\begin{figure}
\begin{centering}
\begin{tabular}{ll}
a) & b)\tabularnewline
\includegraphics[width=8cm]{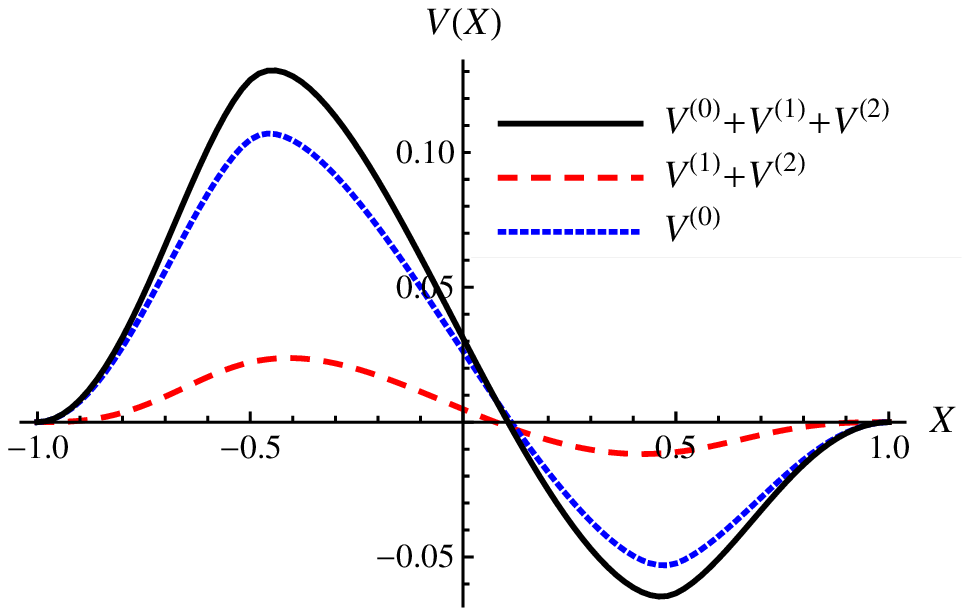} &
\includegraphics[width=8cm]{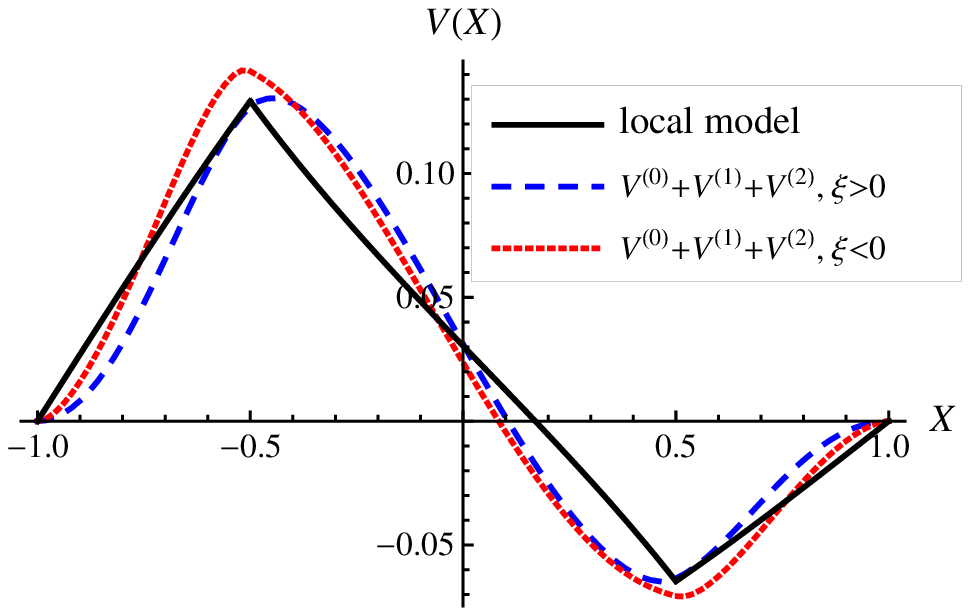}
\tabularnewline
\end{tabular}
\par\end{centering}
\caption{Vector TDA for $M=350\,\mathrm{MeV}$, $t=-0.1\,\mathrm{GeV}^{2}$ and
$n=1$. a) Solid line representing full VTDA for $\xi=0.5$ is the sum of the
part coming from the local part of the vector vertex (dotted) and the
non-local one (dashed); b) comparison of the results of this paper for
$\xi=0.5$ (dashed), $\xi=-0.5$ (dotted) and the local version of the model
$M\left( k\right) =M$ for $\xi=\pm0.5$.}%
\label{fig:vector}%
\end{figure}

Our results are shown in fig. \ref{fig:vector}. We plot also separately the
local part $V^{\left(  0\right)  }$ and the addition $V^{\left(  1\right)
}+V^{\left(  2\right)  }$ itself. We compare the new result with the local
version of our model calculated in \cite{KotkoMP_TDA}. Qualitative behavior
with $M,$ $\xi,$ $n$ and $t$ is the same as in \cite{KotkoMP_TDA} therefore
we do not discuss this further. However now the condition coming from axial
anomaly
\begin{equation}
\int_{-1}^{1}V\left(  X,\xi,t\right)  =\frac{1}{2\pi^{2}}\label{eq:6}%
\end{equation}
is satisfied automatically for any value of $n$ and $\Lambda$ (which does not
have to be necessarily equal $\Lambda_{n}$).

\subsection{Axial TDA}

\label{sub:ATDA}

For the matrix element of the axial operator we have%
\begin{multline}
\int\frac{d\lambda}{2\pi}e^{i\lambda Xp^{+}} 
\left\langle \gamma\left(  P_{2},\varepsilon\right)  \left\vert
\overline{d}\left(  -\frac{\lambda}{2}n\right)  \Gamma_{5}^{\mu}\, u\left(
\frac{\lambda}{2}n\right)  \right\vert \pi^{+}\left(  P_{1}\right)
\right\rangle \\
 =-\frac{\sqrt{2}eMN_{c}}{F_{\pi}} \left(  Q_{d}\mathcal{M}_{A}^{\mu\nu
}\left(  X,\xi,t\right)  -Q_{u}\mathcal{M}_{A}^{\mu\nu}\left(  -X,\xi
,t\right)  \right)  \varepsilon_{\nu}^{\ast},\label{eq:A1}%
\end{multline}
where
\begin{multline}
\mathcal{M}_{A}^{\mu\nu}=\int\frac{d^{4}k}{\left(  2\pi\right)  ^{4}}%
\,\delta\left(  k^{+}-\left(  X-1\right)  p^{+}\right)  \times\nonumber\\
\mathrm{Tr}\left\{  S\left(  k+P_{2}\right)  \Gamma_{5}^{\mu}\left(
k+P_{2},k+P_{1}\right)  S\left(  k+P_{1}\right)  \Gamma_{5}\left(
k+P_{1},k\right)  S\left(  k\right)  \Gamma^{\nu}\left(  k+P_{1},k\right)
\right\}  .\label{eq:A2}%
\end{multline}
This expression looks very similar to \eqref{eq:3}, however here full axial
vertex $\Gamma_{5}^{\mu}$ given in \eqref{eq:axial_vertex} appears instead of
the vector one.

Exploring this expression we find that it has again four parts, similarly to
the vector case. However only two of them contribute to the ATDA, namely:
\begin{equation}
\mathcal{M}_{A}^{\mu\nu,\,\left(  0\right)  }=\int\hat{dk}\,\mathrm{Tr}%
\left\{  S\left(  k+P_{2}\right)  \gamma^{\mu}\gamma_{5}S\left(
k+P_{1}\right)  \Gamma_{5}\left(  k+P_{1},k\right)  S\left(  k\right)
\gamma^{\nu}\right\} \label{eq:A3}%
\end{equation}
and%
\begin{equation}
\mathcal{M}_{A}^{\mu\nu,\,\left(  1\right)  }=\int\hat{dk}\,\mathrm{Tr}%
\left\{  S\left(  k+P_{2}\right)  \gamma^{\mu}\gamma_{5}S\left(
k+P_{1}\right)  \Gamma_{5}\left(  k+P_{1},k\right)  S\left(  k\right)  g^{\nu
}\left(  k+P_{1},k\right)  \right\} \label{eq:A4}%
\end{equation}
The remaining terms are connected with pion DA (or they are gauge artifacts).
This can be seen by noting that%
\begin{equation}
g_{5}^{\mu}\left(  k+P_{2},k+P_{1}\right)  =\frac{q^{\mu}}{t}\,\left(
M\left(  k+P_{2}\right)  +M\left(  k+P_{1}\right)  \right)  \gamma
_{5}\label{eq:A5}%
\end{equation}
and comparing with definition \eqref{eq:defATDA}. Therefore we have%
\begin{equation}
A\left(  X,\xi,t\right)  =A^{\left(  0\right)  }\left(  X,\xi,t\right)
+A^{\left(  1\right)  }\left(  X,\xi,t\right)  ,\label{eq:A6}%
\end{equation}
where again $A^{\left(  0\right)  }$ is the old part already calculated in
\cite{KotkoMP_TDA}. All further details concerning calculations are relegated
to Appendix \ref{sub:App_Axial}.

\begin{figure}[ptb]
\begin{centering}
\begin{tabular}{ll}
a) & b)\tabularnewline
\includegraphics[width=8cm]{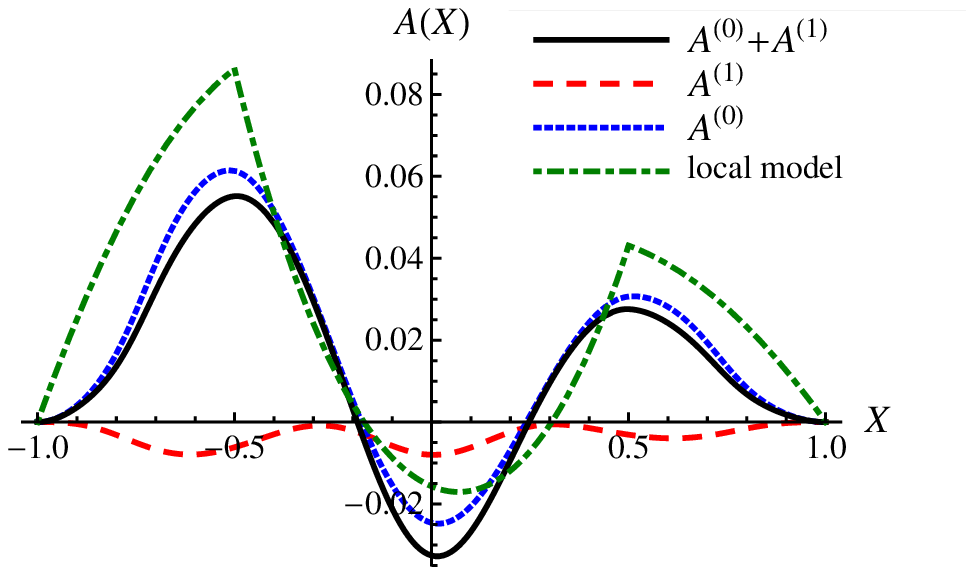} & \includegraphics[width=8cm]{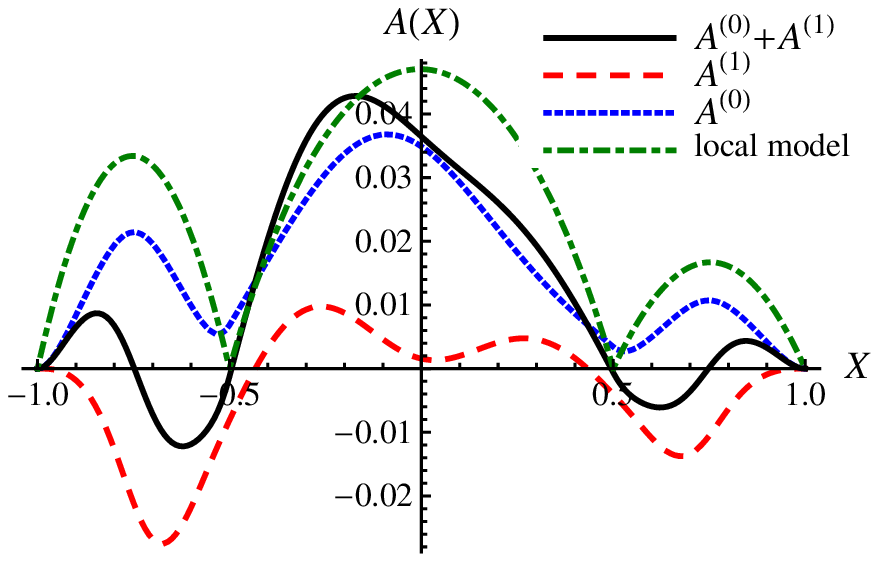}\tabularnewline
c) & d)\tabularnewline
\includegraphics[width=8cm]{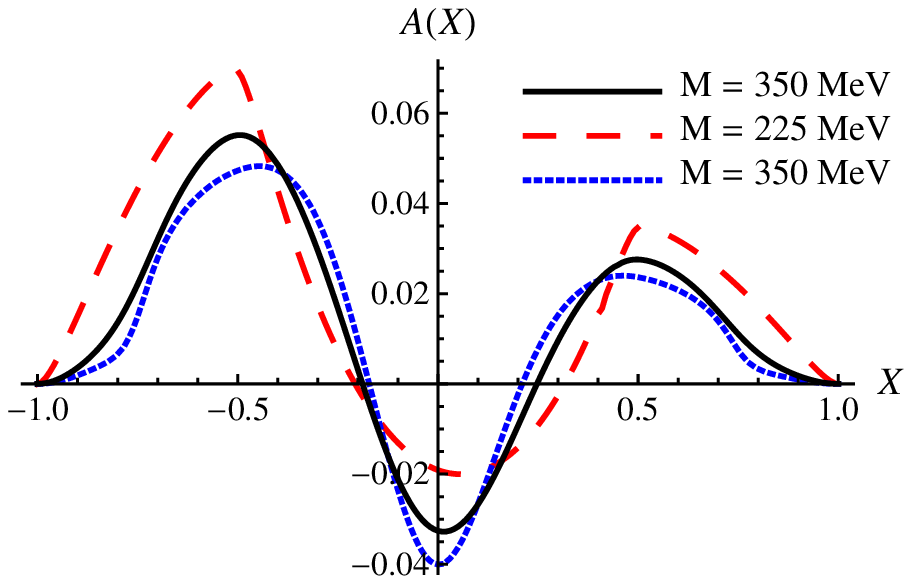} & \includegraphics[width=8cm]{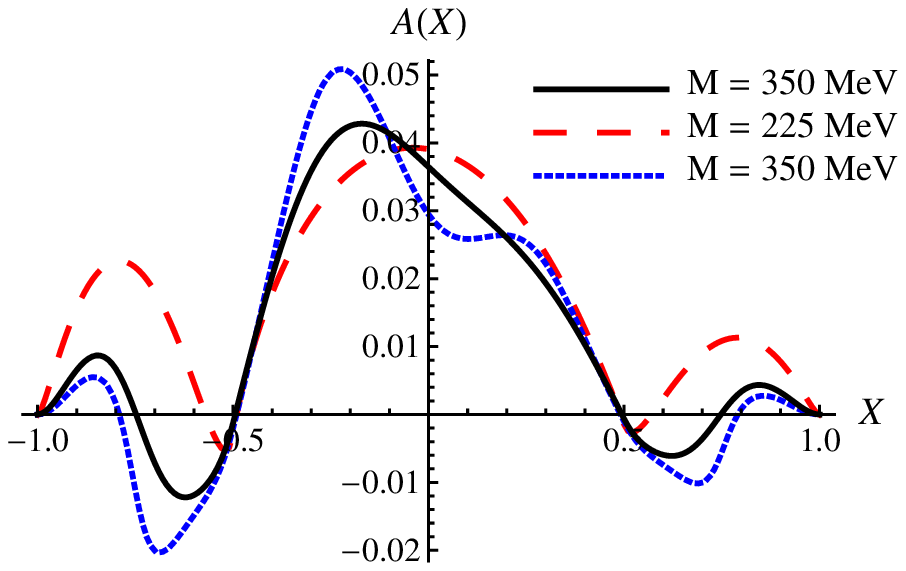}\tabularnewline
\end{tabular}
\par\end{centering}
\caption{Axial TDA for $M=350\,\mathrm{MeV}$, $t=-0.1\,\mathrm{GeV}^{2}$ and
$n=1$. a) Solid line representing full ATDA for $\xi=0.5$ is the sum of the
part coming from the local part of the vector vertex (dotted) and the
non-local one (dashed). Dash-dotted curve is the results for the local model
$M\left( k\right) =M$. b) The same for $\xi=-0.5$; c) results for positive
$\xi=0.5$, $n=1$, $t=-0.1\,\mathrm{GeV}^{2}$ and different values of $M$; d)
the same for negative $\xi=-0.5$.}%
\label{fig:axial}%
\end{figure}

The results are shown in fig. \ref{fig:axial}. It turns out that the integral
over $dX$ of $A^{\left( 1\right) }\left( X,\xi,t\right) $ part is negative,
what shifts the value of the axial form factor towards the experimental value.
For negative values of $\xi$ the addition $A^{\left( 1\right) }\left(
X,\xi,t\right) $ has in a sense unexpected shape, what quite drastically
changes the shape of full ATDA for $\xi<0$. This effect is weaker for lower
constituent quark masses as the addition $A^{\left( 1\right) }$ is
proportional to the third power of $M/\Lambda_{n}$. We have checked that for
any $\xi$ (both positive and negative)
\begin{equation}
\int_{-1}^{1}dX\, A^{\left( 1\right) }\left( X,\left| \xi\right| ,t\right)
=\int_{-1}^{1}dX\, A^{\left( 1\right) }\left( X,-\left| \xi\right| ,t\right)
=f\left( t\right) <0,\label{eq:A7}%
\end{equation}
where $f$ is some function of $t$ only. Equation \eqref{eq:A7} is a special
case of the polynomiality condition, which is satisfied in our model, both for
VTDA and ATDA.

\section{Form factors}

\label{sec:formfactors}

\begin{figure}[ptb]
\begin{centering}
\includegraphics[width=12cm]{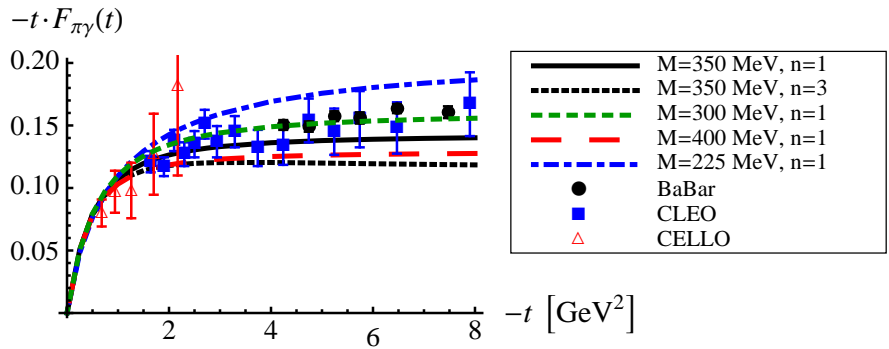}
\par\end{centering}
\caption{Pion to photon transition form factor times momentum transfer for
several model parameters versus CELLO, CLEO and BaBar data in the
$0-8\,\mathrm{GeV}^{2}$ range.}%
\label{fig:ffactor}%
\end{figure}

As already mentioned in Section \ref{sub:VTDA} vector and transition form
factors at zero momentum transfer are reproduced correctly, as given by axial
anomaly. Full result for transition form factor multiplied by momentum
transfer is shown in fig. \ref{fig:ffactor}. We make a comparison with CELLO
\cite{CELLO}, CLEO \cite{CLEO} and some new BaBar data \cite{BABAR} for
several values of model parameters. Since our approach is reliable at low
momenta transfer we limited ourselves to the region of a few $\mathrm{GeV}%
^{2}$. We do not use the standard dipol parametrization for the transition
form factor \cite{CLEO} as a reference, because recent BaBar data seem to
reveal different behavior (see also \cite{Radyushkin,PolyakovDA}). We find that model
parameters $M=300\,\mathrm{MeV}$ and $n=1$ fits the data quite well, while low
constituent quark masses lie much below. However, we underline that momentum
transfer around a few $\mathrm{GeV}^{2}$ can be too large to be treated within
the considered approach.

\begin{center}
\begin{table}[ptb]
\begin{centering}
\begin{tabular}{|c|c|c|c|}
\hline
$M\,\left[\mathrm{MeV}\right]$  & $n$  & $F_{A}\left(0\right)$  & $F_{A}\left(0\right)/F_{V}\left(0\right)$\tabularnewline
\hline
\hline
225  & 1  & 0.0217  & 0.80\tabularnewline
\hline
350  & 1  & 0.0168  & 0.62\tabularnewline
\hline
350  & 5  & 0.0163  & 0.60\tabularnewline
\hline
400  & 1  & 0.0161  & 0.60\tabularnewline
\hline
400  & 5  & 0.0152  & 0.56\tabularnewline
\hline
\end{tabular}
\par\end{centering}
\caption{Numerical values for axial form factor at zero momentum transfer
$F_{A}\left( 0\right) $ for different model parameters.}%
\label{tab:FA}%
\end{table}
\end{center}

In the case of the axial form factor, we find that due to the contribution of
$A^{(1)}$ its value at zero momentum transfer $F_{A}\left(  0\right)  $ is
lower than for the vector one. The values of $F_{A}\left(  0\right)  $ and the
relevant ratios $F_{A}\left(  0\right)  /F_{V}\left(  0\right)  $ for several
model parameters are presented in Table \ref{tab:FA}. We notice that when $n$
is increasing, the value of $F_{A}\left(  0\right)  $ is decreasing very
slowly. Therefore we conclude that $F_{A}\left(  0\right)  $ is quite robust
as far as model parameters are concerned. Experimental value given by PDG is
\begin{equation}
F_{A}^{\mathrm{exp}}\left(  0\right)  =0.0115\pm0.0005\label{eq:F1}%
\end{equation}
and%
\begin{equation}
\left(  F_{A}\left(  0\right)  /F_{V}\left(  0\right)  \right)  _{\mathrm{exp}%
}=0.7_{-0.2}^{+0.6}.\label{eq:F2}%
\end{equation}
Although our predictions are much better then the ones obtained in local
models (with $F_{A}\left(  0\right)  =F_{V}\left(  0\right)  $)
\cite{Tiburzi,BronTDA,Noguera,KotkoMP_TDA}, it still does not agree with
experimental value. Notice however that also prediction for $F_{V}\left(
0\right)  $ provided by conservation of vector current and axial anomaly%
\begin{equation}
F_{V}\left(  0\right)  \approx0.027\label{eq:F3}%
\end{equation}
overshoots the experimental value (PDG)%
\begin{equation}
F_{V}^{\mathrm{exp}}\left(  0\right)  =0.017\pm0.008.\label{eq:F4}%
\end{equation}
We recall also that our calculations were done in the chiral limit. The
effects due to the finite pion mass were discussed in \cite{Noguera} within
local NJL model and turn out to be small.

\section{Summary}

\label{sec:summary}

In the present paper w have used the instanton motivated non-local chiral quark
model to compute pion-to-photon transition distributions
amplitudes in vector and axial channel defined in eqs.(\ref{eq:def_VTDA}) and
(\ref{eq:defATDA}). Our approach follows closely the approach of ref.
\cite{KotkoMP_TDA} with one important modification. In ref.\cite{KotkoMP_TDA}
we have used the momentum dependent constituent quark mass and the nonlocal
pion-quark interaction, however, both vector and axial currents were not
modified to satisfy Ward-Takahashi identities. We have argued in
\cite{KotkoMP_TDA} that such an approach -- although inconsistent -- gives
reasonable shape for the TDA's missing, however, normalization. Indeed, our
present findings confirm this conclusion.

In order to satisfy WT identities the "naive" QCD currents require
modifications if the quark mass depends on momentum. Such modifications are
not unique because WT identities do not fix the longitudinal part of the
current. In the present paper we have used the simplest possible modifications
of the vector and axial currents given by eqs.(\ref{eq:vector_vertex}) and
(\ref{eq:axial_vertex}).

More importantly, we have used the nonlocal vertex not only for the photon
coupling to the quark loop but also for the vector operator defining the TDA.
This modification is not obvious for the following reason. Factorization of
the physical process into hard and soft parts is done in QCD in terms of the
operator product expansion and the relevant operators entering the definition
of the soft part are therefore indeed the "naive" QCD operators. In our
language this would correspond to neglecting $\mathcal{M}_{V}^{\mu
\nu,\,\left(  2\right)  }$ of eq. (\ref{eq:V4}). We have checked that this
would lead to the violation of the normalization condition for VTDA. In the
case of ATDA the non-local part of the axial current does not contribute,
since it enters the piece proportional to the pion DA -- see
eq.(\ref{eq:defATDA}) -- but not the ATDA itself.

Our findings can be summarized as follows. The normalization of TDA's in the
vector channel -- which is fixed by the axial anomaly -- turns out to be
correctly reproduced in our approach. At the same time normalization of ATDA's
is lowered due to the non-local term in the vector current. As a result the
two normalizations are no longer equal, as observed experimentally. The shapes
of the TDA's are similar to the ones obtained in the local model and in the
non-local model with naive currents. In the case of VTDA's we observe,
however, small difference between distributions with positive and negative
$\xi$ what was previously not the case. ATDA's acquire negative contribution
from the non-local piece which makes the differences between positive and
negative $\xi$'s even more profound than in the previously considered models.
We have also investigated dependence of TDA's on the choice of constituent
quark mass $M$ and the shape of the cutoff function $F(k)$. Dependence on the
power $n$ entering $F(k)$ is rather weak, whereas the dependence on $M$ is
effectively important only for ATDA for negative $\xi$.

We have also calculated the corresponding form factors. Axial form factor is
quite robust with respect to the shape of the mass dependence on momentum.
Pion-to-photon transition form factor, which is directly related to the vector
form factor, can be compared with experimental data. We compare it with CELLO,
CLEO and recent BaBar data in low energy regime in fig.(\ref{fig:ffactor}) and
conclude that our results reproduce quite well its $t$ dependence.

Let us finish by a remark that we have worked in the chiral limit where
$m_{\pi}=0$. The effects of adding small current masses are expected to be
small, however, only further study can show whether they will be able to
correct small deviations from experimental values of $F_{V,A}(0)$. Finally the
modifications of the nonlocalities that are still possible as far as
longitudinal part of the currents is concerned deserve further investigation.

\begin{acknowledgments}
The authors are grateful to W. Broniowski, E. Ruiz Arriola and A. Dorokhov for
discussions. The paper was partially supported by the Polish-German
cooperation agreement between Polish Academy of Science and DFG.
\end{acknowledgments}

\appendix


\section{Vector TDA}

\label{sub:App_Vector}

In the beginning we introduce a shorthand notation in order to make the
formulae more compact: following abbreviation%
\begin{equation}
M\left(  k\right)  \equiv M_{k}\label{eq:Not1}%
\end{equation}
and similarly for other $k$-dependent quantities. We recall that
$M_{k}=M\,F_{k}^{2}$. For the inverse scalar propagators we use%
\begin{equation}
D_{k}=k^{2}-M_{k}^{2}.\label{eq:Not2}%
\end{equation}

Let us now turn to VTDA. We obtained the following expressions for subsequent
pieces giving contribution to $V\left(  X,\xi,t\right)  $%
\begin{equation}
V^{\left(  i\right)  }\left(  X,\xi,t\right)  =16i\,MN_{c}p^{+}\,\left(
Q_{d}\mathcal{I}^{\left(  i\right)  }\left(  X,\xi,t\right)  +Q_{u}%
\mathcal{I}^{\left(  i\right)  }\left(  -X,\xi,t\right)  \right)
\label{eq:V1app}%
\end{equation}
where%
\begin{equation}
\mathcal{I}^{\left(  0\right)  }\left(  X,\xi,t\right)  =\int\hat{dk}%
\,\frac{F_{k}F_{k+P_{1}}}{D_{k+P_{1}}D_{k+P_{2}}D_{k}}
\times\bigg\{F_{1}\left(  M_{k+P_{1}}-M_{k+P_{2}}\right)  +F_{2}\left(
M_{k}-M_{k+P_{2}}\right)  +M_{k}\bigg\},\label{eq:V2app}%
\end{equation}%
\begin{equation}
\mathcal{I}^{\left(  1\right)  }\left(  X,\xi,t\right)  =\int\hat{dk}%
\,\frac{\left(  -k_{T}^{2}\sin^{2}\theta_{T}\right)  \,F_{k}F_{k+P_{1}}\left(
M_{k}-M_{k+P_{2}}\right)  }{D_{k+P_{1}}D_{k+P_{2}}D_{k}\,(k\cdot P_{2}%
)},\label{eq:V3app}%
\end{equation}%
\begin{equation}
\mathcal{I}^{\left(  2\right)  }\left(  X,\xi,t\right)  =-\int\hat{dk}%
\,\frac{\left(  -k_{T}^{2}\sin^{2}\theta_{T}\right)  \,F_{k}F_{k+P_{1}}\left(
M_{k+P_{2}}-M_{k+P_{1}}\right)  }{D_{k+P_{1}}D_{k+P_{2}}D_{k}\,(k\cdot
q)}.\label{eq:V4app}%
\end{equation}
In the above formulae%
\begin{equation}
F_{1}=-\frac{\vec{k}_{T}\cdot\vec{q}_{T}}{t\left(  1+\xi\right)  }-\frac{1}%
{2}\left(  X-1\right)  ,\label{eq:V5}%
\end{equation}%
\begin{equation}
F_{2}=2\xi\frac{\vec{k}_{T}\cdot\vec{q}_{T}}{t\left(  \xi^{2}-1\right)
}+\left(  X-1\right)  \label{eq:V6}%
\end{equation}
and $\theta_{T}$ is an angle between $\vec{k}_{T}$ and $\vec{q}_{T}$ in the
transverse plane. Integration measure $\hat{dk}$ was introduced in eq.
(\ref{eq:V1a}).

As an example we briefly explain how to calculate $\mathcal{I}^{\left(
1\right)  }$. Following refs. \cite{MP_pion} we introduce dimensionless
variables $\kappa=k/\Lambda,$ $\bar{P}_{1}=P_{1}/\Lambda,$ $\bar{P}_{2}%
=P_{2}/\Lambda,$ $r=M/\Lambda$ and use
\begin{align}
u_{1,2} &  =\left(  \kappa+\bar{P}_{1,2}\right)  ^{2}-1+i\epsilon
,\label{eq:V8}\\
u_{3} &  =\kappa^{2}-1+i\epsilon.\label{eq:V9}%
\end{align}
Then using \eqref{Fkdef} we have%
\begin{equation}
\mathcal{I}^{\left(  1\right)  }\left(  X,\xi,t\right)  =-\frac{r}{\Lambda
^{2}}\int\hat{d\kappa}\,\frac{\kappa_{T}^{2}\sin^{2}\theta_{T}\,u_{1}%
^{3n}u_{2}^{2n}u_{3}^{n}\left(  u_{2}^{2n}-u_{3}^{2n}\right)  }{(\kappa
\cdot\bar{P_{2}})\,G\left(  u_{1}\right)  G\left(  u_{2}\right)  G\left(
u_{3}\right)  }, \label{eq:X1}
\end{equation}
where%
\begin{equation}
G\left(  z\right)  =z^{4n+1}+z^{4n}-r^{2}.\label{eq:V11}%
\end{equation}
We can obviously write
\begin{equation}
G\left(  z\right)  =\prod_{i=1}^{4n+1}\left(  z-\eta_{i}\right)
,\label{eq:V12}%
\end{equation}
where $\eta_{i}$-s are the roots of $G\left(  z\right)  =0$ to be obtained
numerically. Their properties were discussed in \cite{MP_pion} and
\cite{KotkoMP_foton}. Then using decomposition into proper fractions we
obtain%
\begin{equation}
\mathcal{I}^{\left(  1\right)  }\left(  X,\xi,t\right)  =-\frac{r}{\Lambda
^{2}}\sum_{i,j,k}^{4n+1}f_{i}f_{j}f_{k}\alpha_{ijk}\tilde{\mathcal{I}}%
_{ijk}^{\left(  1\right)  }\left(  X,\xi,t\right),\label{eq:X2}
\end{equation}
where%
\begin{equation}
f_{i}=\prod_{j\neq i}^{4n+1}\frac{1}{\left(  \eta_{i}-\eta_{j}\right)
},\label{eq:V15}%
\end{equation}%
\begin{equation}
\alpha_{ijk}=\eta_{i}^{3n}\eta_{j}^{2n}\eta_{k}^{n}\left(  \eta_{j}^{2n}%
-\eta_{k}^{2n}\right)
\end{equation}
and%
\begin{equation}
\tilde{\mathcal{I}}_{ijk}^{\left(  1\right)  }\left(  X,\xi,t\right)
=\int\hat{d\kappa}\,\frac{\kappa_{T}^{2}\sin^{2}\theta_{T}}{\kappa\cdot
\bar{P_{2}}\left(  u_{1}-\eta_{i}\right)  \left(  u_{2}-\eta_{j}\right)
\left(  u_{3}-\eta_{k}\right)  }.
\end{equation}
Next we assume that this expression can be continued to the Euclidean space --
this is equivalent in Minkowski space to deformation of the $d\kappa^{-}$
integration contour as described in \cite{MP_pion,KotkoMP_foton}. This choice
of the contour assures that the results are real and analytical in
$\Lambda_{n}$. The remaining parts of integrals $\mathcal{I}^{\left(
0\right)  },\,\mathcal{I}^{\left(  2\right)  }$ can be transformed to a sum of
the integrals similarly as in the case $\mathcal{I}^{\left(  1\right)  }$. It
is convenient to start with the $d\kappa^{-}$ integration using residue
theorem. Positions of the poles in the $\kappa^{-}$ complex plane depend on
the kinematical region and on the sign of $\xi$, however the pole connected
with $\kappa\cdot\bar{P}_{2}$ does not give contribution as it should be.
Physical support $-1<X<1$ splits into three pieces: $-1<X<-\left\vert
\xi\right\vert $, $-\left\vert \xi\right\vert <X<\left\vert \xi\right\vert $
and $\left\vert \xi\right\vert <X<1$. Integrals $\mathcal{I}^{\left(
i\right)  }\left(  X,\xi,t\right)  $ are nonzero in the last two, while
$\mathcal{I}^{\left(  i\right)  }\left(  -X,\xi,t\right)  $ in the first two intervals.

In order to write down the results it is useful to introduce the following notation%

\begin{equation}
g_{i}^{(A)}=\left( \kappa_{T}^{2}+1\right) \left( 1-\xi\right) -\vec{\kappa}%
_{T}\cdot\vec{q}_{T}\left( X+1\right) -\left( X-1\right) \left( 1-3\xi-X\left(
1+\xi\right) \right) \bar{p}^{2}+\left( 1-\xi\right) \eta_{i},
\end{equation}
\begin{equation}
g_{i}^{(B)}=\left( \kappa_{T}^{2}+1\right) \left( 1-\xi\right) -\vec{\kappa}%
_{T}\cdot\vec{q}_{T}\left( X-1\right) -\left( X-1\right) ^{2}\left(
1+\xi\right) \bar{p}^{2}+\left( 1-\xi\right) \eta_{i},
\end{equation}
\begin{equation}
h_{i}^{(A)}=\left( \kappa_{T}^{2}+1\right) \xi+X\,\vec{\kappa}_{T}\cdot\vec
{q}_{T}-\left( X^{2}-1\right) \xi\bar{p}^{2}+\xi\eta_{i},
\end{equation}
\begin{equation}
h_{i}^{(B)}=\left( \kappa_{T}^{2}+1\right) \xi+\left( X-1\right) \,\vec{\kappa
}_{T}\cdot\vec{q}_{T}-\left( X-1\right) ^{2}\xi\bar{p}^{2}+\xi\eta_{i},
\end{equation}
\begin{equation}
d_{ij}^{(AB)}=-2\xi\left( \kappa_{T}^{2}+1\right) -2X\,\vec{\kappa}_{T}\cdot
\vec{q}_{T}+2\left( X^{2}-1\right) \xi\bar{p}^{2}+\left( X-\xi\right) \eta
_{i}-\left( X+\xi\right) \eta_{j},
\end{equation}
\begin{equation}
d_{ij}^{\left( AC \right) }=-\left( 1+\xi\right) \left( \kappa_{T}^{2}+1\right) -\left(
1-X\right) \,\vec{\kappa}_{T}\cdot\vec{q}_{T}+\left( 1-\xi\right) \left(
X-1\right) ^{2}\bar{p}^{2}+\left( X+\xi\right) \eta_{i}-\left( X-1\right)
\eta_{j},
\end{equation}
\begin{equation}
d_{ij}^{(CB)}=\left( 1-\xi\right) \left( \kappa_{T}^{2}+1\right) +\left(
1-X\right) \,\vec{\kappa}_{T}\cdot\vec{q}_{T}+\left( 1+\xi\right) \left(
X-1\right) ^{2}\bar{p}^{2}+\left( X-\xi\right) \eta_{i}-\left( X-1\right)
\eta_{j},
\end{equation}
\begin{equation}
\beta_{ijk}=-\alpha_{kji}
\end{equation}
where $\bar{p}=p/\Lambda$ is dimensionless average momentum. Then we have

\begin{itemize}
\item $\left| \xi\right| <X<1$%

\begin{equation}
\mathcal{I}^{\left( 0\right) }\left( X,\left| \xi\right| ,t\right)
=\mathcal{I}^{\left( 0\right) }\left( X,-\left| \xi\right| ,t\right) =\frac
{1}{2}\frac{i\, r}{\Lambda p^{+}}\,\left( X-1\right) 
\sum_{i,j,k}^{4n+1}f_{i}f_{j}f_{k}\,\int\frac{d^{2}\kappa_{T}}{\left(
2\pi\right) ^{3}}\,\frac{\alpha_{ijk}F_{2}+\alpha_{kji}F_{1}+\eta_{i}^{3n}%
\eta_{j}^{4n}\eta_{k}^{n}}{d_{ki}^{(AC)}d_{kj}^{(CB)}}%
\end{equation}
\begin{equation}
\mathcal{I}^{\left( 1\right) }\left( X,\left| \xi\right| ,t\right)
=\mathcal{I}^{\left( 1\right) }\left( X,-\left| \xi\right| ,t\right)
=\frac{i\, r}{\Lambda p^{+}}\,\left( X-1\right) ^{2}\sum_{i,j,k}^{4n+1}%
\alpha_{ijk}f_{i}f_{j}f_{k}\,\int\frac{d^{2}\kappa_{T}}{\left( 2\pi\right)
^{3}}\,\frac{-\kappa_{T}^{2}\sin^{2}\theta_{T}}{g_{k}^{(B)}d_{ki}^{(AC)}%
d_{kj}^{(CB)}}%
\end{equation}
\begin{equation}
\mathcal{I}^{\left( 2\right) }\left( X,\left| \xi\right| ,t\right)
=\mathcal{I}^{\left( 2\right) }\left( X,-\left| \xi\right| ,t\right) =\frac
{1}{2}\frac{i\, r}{\Lambda p^{+}}\,\left( X-1\right) ^{2}\sum_{i,j,k}%
^{4n+1}\beta_{ijk}f_{i}f_{j}f_{k}\,\int\frac{d^{2}\kappa_{T}}{\left(
2\pi\right) ^{3}}\,\frac{-\kappa_{T}^{2}\sin^{2}\theta_{T}}{h_{k}^{(B)}%
d_{ki}^{(AC)}d_{kj}^{(CB)}}%
\end{equation}

\item $-\left| \xi\right| <X<\left| \xi\right| $%

\begin{equation}
\mathcal{I}^{\left( 0\right) }\left( X,\left| \xi\right| ,t\right) =-\frac
{1}{2}\frac{i\, r}{\Lambda p^{+}}\,\left( X+\xi\right) \sum_{i,j,k}%
^{4n+1}f_{i}f_{j}f_{k}\,\int\frac{d^{2}\kappa_{T}}{\left( 2\pi\right) ^{3}%
}\,\frac{\alpha_{ijk}F_{2}+\alpha_{kji}F_{1}+\eta_{i}^{3n}\eta_{j}^{4n}%
\eta_{k}^{n}}{d_{ij}^{(AB)}\left( -d_{ki}^{(AC)}\right) }%
\end{equation}
\begin{equation}
\mathcal{I}^{\left( 0\right) }\left( X,-\left| \xi\right| ,t\right) =-\frac
{1}{2}\frac{i\, r}{\Lambda p^{+}}\,\left( X-\xi\right) \sum_{i,j,k}%
^{4n+1}f_{i}f_{j}f_{k}\,\int\frac{d^{2}\kappa_{T}}{\left( 2\pi\right) ^{3}%
}\,\frac{\alpha_{ijk}F_{2}+\alpha_{kji}F_{1}+\eta_{i}^{3n}\eta_{j}^{4n}%
\eta_{k}^{n}}{d_{ij}^{(AB)}d_{kj}^{(CB)}}%
\end{equation}
\begin{equation}
\mathcal{I}^{\left( 1\right) }\left( X,\left| \xi\right| ,t\right) =-\frac{i\,
r}{\Lambda p^{+}}\,\left( X+\xi\right) ^{2}\sum_{i,j,k}^{4n+1}\alpha
_{ijk}f_{i}f_{j}f_{k}\,\int\frac{d^{2}\kappa_{T}}{\left( 2\pi\right) ^{3}%
}\,\frac{-\kappa_{T}^{2}\sin^{2}\theta_{T}}{g_{i}^{(A)}d_{ij}^{(AB)}\left(
-d_{ki}^{(AC)}\right) }%
\end{equation}
\begin{equation}
\mathcal{I}^{\left( 1\right) }\left( X,-\left| \xi\right| ,t\right)
=-\frac{i\, r}{\Lambda p^{+}}\,\left( X-\xi\right) ^{2}\sum_{i,j,k}%
^{4n+1}\alpha_{ijk}f_{i}f_{j}f_{k}\,\int\frac{d^{2}\kappa_{T}}{\left(
2\pi\right) ^{3}}\,\frac{-\kappa_{T}^{2}\sin^{2}\theta_{T}}{g_{j}^{(B)}%
d_{ij}^{(AB)}d_{kj}^{(CB)}}%
\end{equation}
\begin{equation}
\mathcal{I}^{\left( 2\right) }\left( X,\left| \xi\right| ,t\right) =-\frac
{1}{2}\frac{i\, r}{\Lambda p^{+}}\,\left( X+\xi\right) ^{2}\sum_{i,j,k}%
^{4n+1}\beta_{ijk}f_{i}f_{j}f_{k}\,\int\frac{d^{2}\kappa_{T}}{\left(
2\pi\right) ^{3}}\,\frac{-\kappa_{T}^{2}\sin^{2}\theta_{T}}{h_{i}^{(A)}%
d_{ij}^{(AB)}\left( -d_{ki}^{(AC)}\right) }%
\end{equation}
\begin{equation}
\mathcal{I}^{\left( 2\right) }\left( X,-\left| \xi\right| ,t\right) =-\frac
{1}{2}\frac{i\, r}{\Lambda p^{+}}\,\left( X-\xi\right) ^{2}\sum_{i,j,k}%
^{4n+1}\beta_{ijk}f_{i}f_{j}f_{k}\,\int\frac{d^{2}\kappa_{T}}{\left(
2\pi\right) ^{3}}\,\frac{-\kappa_{T}^{2}\sin^{2}\theta_{T}}{h_{i}^{(A)}%
d_{ij}^{(AB)}d_{kj}^{(CB)}}%
\end{equation}

\end{itemize}

Integral $\mathcal{I}^{\left(  0\right)  }$ can be calculated completely
analytically, however we do not present the result here to make the paper more
compact. It should be pointed out that the analytic formulae contain large
sums of complicated complex expressions -- therefore it is usually more
efficient to perform all integrals numerically.

\section{Axial TDA}

\label{sub:App_Axial}

In the axial channel we get%
\begin{equation}
A^{\left( i\right) }\left( X,\xi,t\right) =-16i\, M\, N_{c}\, p^{+}\left(
Q_{d}\mathcal{J}^{\left( i\right) }\left( X,\xi,t\right) -Q_{u}\mathcal{J}%
^{\left( i\right) }\left( -X,\xi,t\right) \right) ,\label{eq:Ax1}%
\end{equation}
where%
\begin{multline}
\mathcal{J}^{\left( 0\right) }\left( X,\xi,t\right) =\int\hat{dk}\,\frac
{F_{k}F_{k+P_{1}}}{D_{k+P_{1}}D_{k+P_{2}}D_{k}}\\
\times\bigg\{F_{1}\big[M_{k+P_{1}}-M_{k+P_{2}}-2M_{k}\big]-F_{2}\left(
M_{k+P_{2}}-M_{k}\right) +M_{k}+2F_{3}\,\big[M_{k+P_{1}}-M_{k}%
\big]\bigg\},\label{eq:Ax}%
\end{multline}
\begin{multline}
\mathcal{J}^{\left( 1\right) }\left( X,\xi,t\right) =\int\hat{dk}\,\frac
{F_{k}F_{k+P_{1}}\left( M_{k}-M_{k+P_{2}}\right) }{k\cdot P_{2}D_{k+P_{1}%
}D_{k+P_{2}}D_{k}}\\
\times\bigg\{F_{1}\big[M_{k+P_{1}}M_{k}-M_{k+P_{2}}M_{k}-2k^{2}-2k\cdot
p\big]+F_{3}\big[M_{k+P_{1}}M_{k+P_{2}}-M_{k}M_{k+P_{2}}+M_{k}M_{k+P_{1}}\\
-k^{2}+P_{1}\cdot P_{2}\big]\bigg\}.\label{eq:Axapp}%
\end{multline}
Functions $F_{1},$ $F_{2}$ are the same as in vector case while%
\begin{equation}
F_{3}=\frac{8\xi\left( \vec{k}_{T}\cdot\vec{q}_{T}\right) ^{2}+2t\,\vec{k}%
_{T}\cdot\vec{q}_{T}\left( X-1\right) \left( 2\xi^{2}+\xi-1\right) +t\,\left(
\xi^{2}-1\right) \left( t\,\left( X-1\right) ^{2}\left( 1+\xi\right) -4\xi
k_{T}^{2}\right) }{2\left( 1-\xi\right) \left( 1+\xi\right) ^{2}t^{2}}.
\end{equation}

Calculations proceed quite similarly as before, although they are more
involved. However, in the case of $\mathcal{J}^{\left(  1\right)  }$
additional difficulty arises if one wants to obtain result for any $n$. Let us
discuss this further.

Problematic part comes from the subterm of $\mathcal{J}^{\left(  1\right)  }$
containing more then $4$ powers of $F(k)$. Using \eqref{eq:V8},\eqref{eq:V9}
this part reads%
\begin{equation}
\mathcal{J}_{A}^{\left(  1\right)  }\left(  X,\xi,t\right)  =\frac{r^{3}%
}{\Lambda^{2}}\int\hat{dk}\,\frac{u_{1}^{n}u_{2}^{2n}\left(  u_{2}^{2n}%
-u_{1}^{2n}\right)  \left(  F_{1}+F_{3}\right)  }{\kappa\cdot\bar{P}%
_{2}G\left(  u_{1}\right)  G\left(  u_{2}\right)  G\left(  u_{3}\right)
u_{3}^{n}}.
\end{equation}
Notice that there is an additional $u_{3}^{n}$ in denominator which gives rise
to multiple poles if $n\geq2$. However if $n=1$
\begin{equation}
\mathcal{J}_{A,\,n=1}^{\left(  1\right)  }\left(  X,\xi,t\right)  =\frac
{r^{3}}{\Lambda^{2}}\sum_{i,j}^{5}\sum_{k}^{6}f_{i}f_{j}\tilde{f}_{k}\int
\hat{dk}\,\frac{\left(  \eta_{i}\eta_{j}^{4}-\eta_{i}^{3}\eta_{j}^{2}\right)
\left(  F_{1}+F_{3}\right)  }{\kappa\cdot\bar{P}_{2}\left(  u_{1}-\eta
_{i}\right)  \left(  u_{2}-\eta_{j}\right)  \left(  u_{3}-\eta_{k}\right)  },
\end{equation}
where%
\begin{equation}
\tilde{f}_{i}=\prod_{j=0,j\neq i}^{5}\frac{1}{\left(  \eta_{i}-\eta
_{j}\right)  }%
\end{equation}
and $\eta_{0}=0$, while $\eta_{1},\ldots,\eta_{5}$ remain solutions of $G=0$.
In order to avoid calculating residues of the multiple poles if $n\geq2$, we
choose to close the contour in such a way that multiple pole does not lie
inside. The price we pay is that relevant expressions become more complicated.
Moreover, although the pole coming from $\kappa\cdot\bar{P}_{2}$ does not give
contribution to $\mathcal{J}^{\left(  1\right)  }$, it does contribute to
$\mathcal{J}_{A}^{\left(  1\right)  }$ and $\mathcal{J}_{B}^{\left(  1\right)
}$ separately where $\mathcal{J}^{\left(  1\right)  }=\mathcal{J}_{A}^{\left(
1\right)  }+\mathcal{J}_{B}^{\left(  1\right)  }$. This fact has to be taken
into account when we decide to choose different contours for $\mathcal{J}%
_{A}^{\left(  1\right)  }$ and $\mathcal{J}_{B}^{\left(  1\right)  }$.

Now we are in a position to write the results. We use the same abbreviations
as in the vector case, additionally we define:%
\begin{equation}
c_{ijk}=-F_{1}\left(  \beta_{ijk}+2\eta_{i}^{3n}\eta_{j}^{4n}\eta_{k}%
^{n}\right)  -2F_{3}\beta_{ijk}+F_{2}\alpha_{ijk}+\eta_{i}^{3n}\eta_{j}%
^{4n}\eta_{k}^{n}%
\end{equation}

\begin{equation}
w_{ijk}^{(A,B,C)}=F_{1}r^{2}\beta_{ijk}-F_{1}\beta_{ijk}t^{(A,B,C)}+F_{3}%
r^{2}\left(  \alpha_{ijk}+\beta_{ijk}\right)  +F_{3}\beta_{ijk}s^{(A,B,C)},
\end{equation}
where%
\begin{equation}
t^{(A)}=\frac{-\left(  1+2\xi\right)  \kappa_{T}^{2}+\left(  1+\eta_{i}%
-\vec{\kappa}_{T}\cdot\vec{q}_{T}\right)  \left(  2X-1\right)  +\left(
X-1\right)  \left(  X\left(  2\xi-1\right)  +1\right)  \bar{p}^{2}}{X+\xi}%
\end{equation}%
\begin{equation}
s^{(A)}=\frac{-\left(  1+\xi\right)  \kappa_{T}^{2}+\left(  1+\eta_{i}%
-\vec{\kappa}_{T}\cdot\vec{q}_{T}\right)  \left(  X-1\right)  -\left(
X-1\right)  \left(  \left(  X-1\right)  ^{2}\left(  1-\xi\right)  +2\left(
X+\xi\right)  \right)  \bar{p}^{2}}{X+\xi}%
\end{equation}%
\begin{equation}
t^{(B)}=\frac{\left(  2\xi-1\right)  \kappa_{T}^{2}+\left(  1+\eta_{j}%
+\vec{\kappa}_{T}\cdot\vec{q}_{T}\right)  \left(  2X-1\right)  -\left(
X-1\right)  \left(  \left(  2X-1\right)  \left(  1+\xi\right)  +X-\xi\right)
\bar{p}^{2}}{X-\xi}%
\end{equation}%
\begin{equation}
s^{(B)}=\frac{\left(  -1+\xi\right)  \kappa_{T}^{2}+\left(  1+\eta_{j}%
+\vec{\kappa}_{T}\cdot\vec{q}_{T}\right)  \left(  X-1\right)  -\left(
X-1\right)  \left(  \left(  X-1\right)  ^{2}\left(  1+\xi\right)  +2\left(
X-\xi\right)  \right)  \bar{p}^{2}}{X-\xi}%
\end{equation}%
\begin{equation}
t^{(C)}=\frac{\kappa_{T}^{2}+\left(  1+\eta_{k}\right)  \left(  2X-1\right)
}{X-1}+\left(  X-1\right)  \bar{p}^{2}%
\end{equation}%
\begin{equation}
s^{(C)}=1+\eta_{k}-2\bar{p}^{2}.
\end{equation}
We recall at this point that $\eta_{0}=0$, therefore although sums below run
from unity, notation like {}\textquotedblleft$d_{i0}$\textquotedblright\ makes sense.

\begin{itemize}
\item $\left| \xi\right| <X<1$%

\begin{equation}
\mathcal{J}^{\left( 0\right) }\left( X,\left| \xi\right| ,t\right)
=\mathcal{J}^{\left( 0\right) }\left( X,-\left| \xi\right| ,t\right)
=\frac{i\, r}{\Lambda p^{+}}\left( X-1\right) \sum_{i,j,k}^{4n+1}f_{i}%
f_{j}f_{k}\,\int\frac{d^{2}\kappa_{T}}{\left( 2\pi\right) ^{3}}\,\frac
{c_{ijk}}{d_{ki}^{(AC)}d_{kj}^{(CB)}}%
\end{equation}
\begin{multline}
\mathcal{J}_{A}^{\left( 1\right) }\left( X,\left| \xi\right| ,t\right)
=\mathcal{J}_{A}^{\left( 1\right) }\left( X,-\left| \xi\right| ,t\right)
=-\frac{i\, r^{3}}{\Lambda p^{+}}\sum_{i,j,k}^{4n+1}f_{i}f_{j}f_{k}\left(
\eta_{i}^{n}\eta_{j}^{4n}-\eta_{i}^{3n}\eta_{j}^{2n}\right) \,\int\frac
{d^{2}\kappa_{T}}{\left( 2\pi\right) ^{3}}\\
\left( F_{1}+F_{3}\right) \bigg\{\frac{\left( X+\xi\right) ^{n+2}}{g_{i}%
^{(A)}d_{ij}^{(AB)}\left( -d_{ki}^{(AC)}\right) \left( -d_{0i}^{(AC)}\right) ^{n}%
}+\frac{\left( X-\xi\right) ^{n+2}}{g_{j}^{(B)}d_{ij}^{(AB)}d_{kj}^{(CB)}\left(
-d_{0j}^{(CB)}\right) ^{n}}+\frac{\left( 1-\xi\right) ^{n+2}}{\left( -g_{i}%
^{(A)}\right) g_{j}^{(B)}g_{k}^{(B)}\left( -g_{0}^{(B)}\right) ^{n}}\bigg\}
\end{multline}
\begin{equation}
\mathcal{J}_{B}^{\left( 1\right) }\left( X,\left| \xi\right| ,t\right)
=\mathcal{J}_{B}^{\left( 1\right) }\left( X,-\left| \xi\right| ,t\right)
=\frac{i\, r}{\Lambda p^{+}}\left( X-1\right) ^{2}\sum_{i,j,k}^{4n+1}%
f_{i}f_{j}f_{k}\,\int\frac{d^{2}\kappa_{T}}{\left( 2\pi\right) ^{3}}%
\,\frac{w_{ijk}^{(C)}}{g_{k}^{(B)}d_{ki}^{(AC)}d_{kj}^{(CB)}}%
\end{equation}

\item $-\left| \xi\right| <X<\left| \xi\right| $%

\begin{equation}
\mathcal{J}^{\left( 0\right) }\left( X,\left| \xi\right| ,t\right) =-\frac{i\,
r}{\Lambda p^{+}}\left( X+\xi\right) \sum_{i,j,k}^{4n+1}f_{i}f_{j}f_{k}%
\,\int\frac{d^{2}\kappa_{T}}{\left( 2\pi\right) ^{3}}\,\frac{c_{ijk}}%
{d_{ki}^{(AC)}d_{kj}^{(CB)}}%
\end{equation}
\begin{equation}
\mathcal{J}^{\left( 0\right) }\left( X,-\left| \xi\right| ,t\right)
=-\frac{i\, r}{\Lambda p^{+}}\left( X-\xi\right) \sum_{i,j,k}^{4n+1}f_{i}%
f_{j}f_{k}\,\int\frac{d^{2}\kappa_{T}}{\left( 2\pi\right) ^{3}}\,\frac
{c_{ijk}}{d_{ij}^{(AB)}d_{kj}^{(CB)}}%
\end{equation}
\begin{equation}
\mathcal{J}_{A}^{\left( 1\right) }\left( X,\left| \xi\right| ,t\right)
=-\frac{i\, r^{3}}{\Lambda p^{+}}\sum_{i,j,k}^{4n+1}f_{i}f_{j}f_{k}\left(
\eta_{i}^{n}\eta_{j}^{4n}-\eta_{i}^{3n}\eta_{j}^{2n}\right) \,\int\frac
{d^{2}\kappa_{T}}{\left( 2\pi\right) ^{3}}\,\frac{\left( F_{1}+F_{3}\right)
\left( X+\xi\right) ^{n+2}}{g_{i}^{(A)}d_{ij}^{(AB)}\left( -d_{ki}^{(AC)}\right)
\left( -d_{0i}^{(AC)}\right) ^{n}}%
\end{equation}
\begin{equation}
\mathcal{J}_{A}^{\left( 1\right) }\left( X,-\left| \xi\right| ,t\right)
=-\frac{i\, r^{3}}{\Lambda p^{+}}\sum_{i,j,k}^{4n+1}f_{i}f_{j}f_{k}\left(
\eta_{i}^{n}\eta_{j}^{4n}-\eta_{i}^{3n}\eta_{j}^{2n}\right) \,\int\frac
{d^{2}\kappa_{T}}{\left( 2\pi\right) ^{3}}\,\frac{\left( F_{1}+F_{3}\right)
\left( X-\xi\right) ^{n+2}}{g_{j}^{(B)}\left( -d_{ij}^{(AB)}\right) \left(
-d_{kj}^{(CB)}\right) \left( -d_{0j}^{(CB)}\right) ^{n}}%
\end{equation}
\begin{equation}
\mathcal{J}_{B}^{\left( 1\right) }\left( X,\left| \xi\right| ,t\right)
=-\frac{i\, r}{\Lambda p^{+}}\left( X+\xi\right) ^{2}\sum_{i,j,k}^{4n+1}%
f_{i}f_{j}f_{k}\,\int\frac{d^{2}\kappa_{T}}{\left( 2\pi\right) ^{3}}%
\,\frac{w_{ijk}^{(A)}}{g_{i}^{(A)}d_{ij}^{(AB)}\left( -d_{ki}^{(AC)}\right) }%
\end{equation}
\begin{equation}
\mathcal{J}_{B}^{\left( 1\right) }\left( X,-\left| \xi\right| ,t\right)
=-\frac{i\, r}{\Lambda p^{+}}\left( X+\xi\right) ^{2}\sum_{i,j,k}^{4n+1}%
f_{i}f_{j}f_{k}\,\int\frac{d^{2}\kappa_{T}}{\left( 2\pi\right) ^{3}}%
\,\frac{w_{ijk}^{(B)}}{g_{j}^{(B)}\left( -d_{ij}^{(AB)}\right) \left( -d_{kj}%
^{(CB)}\right) }%
\end{equation}

\end{itemize}

Notice, that some of the integrals are superficially divergent. Convergence is
assured by the identity%
\begin{equation}
\sum_{i=1}^{4n+1}f_{i}\eta_{i}^{N}=%
\begin{cases}
1 & \mathrm{for}\, N=4n\\
0 & \mathrm{for\, N<4n}%
\end{cases}
,\label{eq:th1}%
\end{equation}
see \cite{KotkoMP_foton} for the proof and discussion.


\end{document}